\documentclass[aps,prd,amsmath,amssymb,amsfonts,floatfix,nofootinbib,superscriptaddress,longbibliography,10pt,bibnotes,reprint,noeprint]{revtex4-2}
\usepackage{amsmath}
\usepackage{amsfonts}
\usepackage{amssymb}
\usepackage{CJKutf8}
\usepackage{verbatim,graphicx}
\usepackage[all]{xy}
\usepackage{simplewick}
\usepackage{setspace}
\usepackage{array}
\usepackage{amsthm}
\usepackage{enumitem}
\usepackage{bbm}
\usepackage{bbold}
\usepackage{mathtools}
\usepackage{subfigure}

\usepackage[english]{babel}
\usepackage[utf8]{inputenc}
\usepackage[T1]{fontenc}

\usepackage{physics}
\usepackage{mathrsfs}
\usepackage{bm}
\usepackage{enumerate}

\usepackage[dvipsnames]{xcolor}

\usepackage[colorlinks,linkcolor=BrickRed,citecolor=MidnightBlue,urlcolor=MidnightBlue]{hyperref}
\def\ii{{\rm i}}

\allowdisplaybreaks[4]

\newcounter{jvcc}
\newcounter{amgg}
\newcounter{jz}

\renewcommand{\tilde}{\widetilde}

\newcommand{\edoc}{\end{document}}
\newcommand{\eref}[1]{(\ref{#1})}
\newcommand{\nn}{\nonumber}

\renewcommand{\i}{\mathrm{i}}

\newcommand{\be}{\begin{eqnarray}}
	\newcommand{\bea}{\begin{eqnarray}}
		\newcommand{\eea}{\end{eqnarray}}
	\newcommand{\beq}{\begin{equation}}
		\newcommand{\ee}{\end{eqnarray}}
	\newcommand{\eeq}{\end{equation}}
\newcommand{\wchi}{\widetilde \chi}

\newcounter{amg}

\begin{document}

\title{Anatomy of information scrambling and decoherence in the integrable Sachdev-Ye-Kitaev model}

\author{Antonio  M. Garc\'\i a-Garc\'\i a}
\email{amgg@sjtu.edu.cn}
\affiliation{Shanghai Center for Complex Physics,
	School of Physics and Astronomy, Shanghai Jiao Tong
	University, Shanghai 200240, China}
\author{\begin{CJK*}{UTF8}{gbsn}
    Chang Liu (刘畅)
\end{CJK*}}
	\email{cl91tp@gmail.com}
	\affiliation{Shanghai Qi Zhi Institute, Shanghai 200232, China}
	\affiliation{Shanghai Center for Complex Physics,
		School of Physics and Astronomy, Shanghai Jiao Tong
		University, Shanghai 200240, China}
	\author{Lucas S\'a}
	\email{ld710@cam.ac.uk}
	\affiliation{TCM Group, Cavendish Laboratory, University of Cambridge, JJ Thomson Avenue, Cambridge CB3 0HE, UK\looseness=-1}
\author{Jacobus J. M. Verbaarschot}
\email{jacobus.verbaarschot@stonybrook.edu}
\affiliation{Center for Nuclear Theory and Department of Physics and Astronomy, Stony Brook University, Stony Brook, New York 11794, USA}
\author{\begin{CJK*}{UTF8}{gbsn}
		Jie-ping Zheng (郑杰平)
\end{CJK*}}
\email{jpzheng@sjtu.edu.cn}
\affiliation{Shanghai Center for Complex Physics,
	School of Physics and Astronomy, Shanghai Jiao Tong
	University, Shanghai 200240, China}

\begin{abstract}
       The growth of information scrambling, captured by out-of-time-order correlation functions (OTOCs), is a central indicator of the nature of many-body quantum dynamics. Here, we compute analytically the complete time dependence of the OTOC for an integrable Sachdev-Ye-Kitaev (SYK) model, $N$ Majoranas with random two-body interactions of infinite range, coupled to a Markovian bath at finite temperature. In the limit of no coupling to the bath, the time evolution of scrambling experiences different stages. For $t \lesssim \sqrt{N}$, after an initial polynomial growth, the OTOC approaches saturation in a power-law fashion with oscillations superimposed.
       At $t \sim \sqrt{N}$, the OTOC reverses trend and starts to decrease linearly in time. The reason for this linear decrease is that, despite being a subleading $1/N$ effect, 
       the OTOC in this region is governed by the spectral form factor of the antisymmetric couplings of the SYK model. The linear decrease stops at $t \sim 2N$, the Heisenberg time, where saturation occurs. The effect of the environment is an overall exponential decay of the OTOC for times longer than the inverse of the coupling strength to the bath.
       The oscillations at $t \lesssim \sqrt{N}$ indicate lack of thermalization---a desired feature for a better performance of quantum information devices.  
\end{abstract}

\maketitle

\section{Introduction}

The growth of quantum uncertainty at different stages of quantum scrambling can be characterized by out-of-time-order correlation functions (OTOCs).
For instance, quantum uncertainty around the Ehrenfest time grows exponentially~\cite{larkin1969,berman1978,jalabert2018} in the semiclassical limit for quantum chaotic systems.
The analytical calculation of OTOCs even for single-particle quantum chaotic systems is quite challenging with relatively few explicit calculations known 
~\cite{larkin1969,jalabert2018,berman1978}.
For integrable systems, the growth is typically power-law. Reported exponential growth~\cite{scaffidi2019,hummel2019,hashimoto2020,cameo2020,chavez2019,tian2022} in certain integrable systems requires \cite{trunin2023,garciamata2023} choosing initial conditions around unstable fixed points so that the state evolution mimics that of quantum chaotic systems for short times. 

For many-body quantum chaotic systems, the Lyapunov exponent was first computed analytically by Kitaev~\cite{kitaev2015} in the low-temperature limit of the SYK model~\cite{kitaev2015,bohigas1971,french1970,french1971,sachdev1993,benet2001,maldacena2016}
consisting of $N$ fermions with $q$-body random interactions of infinite range. For $q > 2$, the dynamics is quantum chaotic at all timescales~\cite{kitaev2015,garcia2016,maldacena2016} with a Lyapunov exponent that saturates a universal bound on chaos~\cite{maldacena2015}. 
Other aspects of the OTOC in the Hermitian SYK model have been studied analytically in Refs.~\cite{maldacena2016,bagrets2017,kitaev2018,cirac2019,gu2022,zhang2023} and numerically, for $q > 2$, in Refs.~\cite{kobrin2020,garcia2023d} where it was necessary to reach $N \geq 50$ Majoranas in order to reproduce the mentioned saturation of the Lyapunov exponent. 
The central role of the OTOC in the description of quantum scrambling has triggered a flurry of activity~\cite{garciamata2023} in different fields: the OTOC has been studied for a random matrix Hamiltonian~\cite{cipolloni2024,torresherrera2017,cotler2017}, for many-body bosonic systems ~\cite{urbina2018,mollabashi2024}, random circuits~\cite{nahum2018,sondhi2018,yoshimura2024b} and Jackiw-Teitelboim gravity~\cite{jackiw1985,teitelboim1983,shenker2015,maldacena2016a,stanford2022}. 

The fate of information scrambling if the Hermiticity condition is relaxed has also attracted a lot of recent interest~\cite{bergamasco2023quantum,yoshida2019disentangling,tuziemski2019out,weinstein2023scrambling,zanardi2021information,huang2020,syzranov2018out,zhai2020a,yoshimura2024a,yoshimura2024b}. For instance, the Lyapunov exponent has been shown~\cite{garcia2024} to vanish at a certain dissipation strength for both a $q > 2$ SYK model coupled to a Markovian bath~\cite{sa2022,kulkarni2022,garcia2022e}
and for a radiative random circuit~\cite{weinstein2023scrambling}. 

Despite these recent  advances, the full time dependence of the OTOC in quantum many-body systems ~\cite{jalabert2018,fortes2019,garciamata2023,mori2024,yoshimura2024b} is still an outstanding problem. 
Here, we address this problem for an integrable ($q = 2$) SYK with Majorana fermions
at finite temperature coupled to a Markovian bath described by the Lindblad formalism~\cite{belavin1969,lindblad1976,gorini1976,breuer2002,manzano2020}. We obtain a compact analytic expression for the OTOC at finite $N$, for all times, and for any value of the coupling to the bath and temperature, which facilitates a detailed description of the relevant timescales and the role of dissipative effects in information scrambling. We note that the calculation of the OTOC in the Hermitian $q=2$ SYK model was discussed for Dirac fermions in Ref.~\cite{gross2017} but was not worked out in detail.

\section{Model and analytic calculation of the OTOC}

We consider an integrable Hermitian SYK Hamiltonian
coupled to a Markovian bath at inverse temperature $\beta$. 
In this case, the dynamics is described by the Lindblad formalism~\cite{breuer2002,lindblad1976} and depends on the choice of jump operators and temperature-dependent couplings. We start our analysis with the infinite-temperature case, $\beta = 0$, where calculations are especially simple. The evolution of the density matrix is governed by the Lindblad equation,
\be \label{eq:densitymsyk2}
\frac{d\rho}{dt} = -i[H,\rho] + \sum_i L_i \rho L_i^\dagger -\frac 12 \sum_i \{ 
L_i^\dagger L_i, \rho \},
\ee
where $H =  i  \sum_{i<j} J_{ij} \chi_i\chi_j$
is the $q = 2$ SYK Hamiltonian with $\{\chi_i,\chi_j\}=\delta_{ij}$, 
$L_i = \sqrt{\mu} \chi_i$, and $J_{ij} = -J_{ij}$ are Gaussian random numbers of zero mean and variance $\langle J_{ij}^2\rangle = J^2/N$. We set $J = 1$, so all results are in units of $J$. The steady state is the identity, namely, a thermofield double state at infinity temperature.
We aim to probe the dynamics by the study of the growth in time of quantum uncertainty represented by the square of anti-commutators,
\begin{equation} \label{eq:ct}
 C(t) = \frac{1}{N(N-1)} \sum_{i\neq j}\langle\{\chi_i(t),\chi_j\}^2\rangle \equiv 2 F(t)+ 2 I(t),
\end{equation} 
  where $\chi_k$ without explicit arguments stands for its value at $t=0$, and  
\be
  F(t) &=&\frac{1} {N(N-1)}\sum_{m\neq n} \langle \chi_m(t) \chi_n \chi_m(t) \chi_n\rangle, 
\nn\\
I(t)&=&C(t)/2 -F(t).
\ee
Here, $F(t)$ is the OTOC, which for quantum chaotic systems captures the exponential growth of the quantum uncertainty around the Ehrenfest time.
The bra-kets denote an average over all states and an average over the random SYK couplings.
We focus on $F(t)$ since the calculation for $I(t)$ is similar.
To determine the $\mu$-dependence of $F(t)$, 
we can restrict ourselves to two specific Majoranas, $\chi_1$ and $\chi_2$, and for $q=2$ one can easily derive the evolution equation~\cite{blocher2019}:
\begin{align}
	\label{otocdef}
	\frac {dF }{dt}=& 2i\langle [H,\chi_1](t) \chi_2 \chi_1(t) \chi_2 \rangle-\mu N \langle \chi_1(t) \chi_2 \chi_1(t) \chi_2\rangle
	\nn\\
	&	-2\mu \sum_n \langle( \chi_n \chi_1 \chi_n) (t) \chi_2 \chi_1(t) \chi_2\rangle.
\end{align}
Note the extra minus sign in the last term [compared to Eq.~(\ref{eq:densitymsyk2})], which arises in the adjoint Lindblad evolution with fermionic jump operators~\cite{PhysRevB.94.155142},
\be
\frac{d\chi_i(t)}{dt} &&= i[H,\chi_i(t)] \nn\\
&&-\mu \sum_n \chi_n^\dagger \chi_i(t)\chi_n   
-\frac \mu2 \sum_n\{ \chi_n^\dagger \chi_n,\chi_i(t)\} 
\ee
when  deriving the evolution equation \eref{otocdef} with $\chi_n^\dagger =\chi_n$. 
Using
\be
\sum_n \chi_n \chi_k\chi_n= - (N/2-1)\chi_k, 
\ee
it is straightforward to show that the $\mu$ dependence of the OTOC factorizes,
\be
F(t,\mu) = e^{-2\mu t} F(t,\mu=0),
\ee
so the $\mu$ dependence on $C(t)$ factorizes as well.

We now turn to the finite temperature case. 
The proof of factorization of the $\mu$ dependence at finite temperature, presented in Appendix~\ref{app:ft}, follows along the same lines as for $\beta = 0$.
 A finite-temperature steady state $\rho = \exp(-\beta H)/Z$, with $Z={\rm Tr}\exp(-\beta H)$ can be prepared by a judicious choice of jump operators with temperature-dependent couplings~\cite{prosen2008} (see Appendix~\ref{app:ft}). In this case, we define the quantum uncertainty as
 \be
   \label{eq:ctfull}
	C(t,\beta,\mu) &=&\sum_{i \neq j} \frac{\langle \{\chi_i(t)\rho^{1/4},\chi_j \rho^{1/4} \}^2\rangle }
	{N(N-1)}
	\nn \\
	&=&e^{-2\mu t}(F(t,\beta)+I(t,\beta))\nn\\
&\equiv& e^{-2\mu t} C(t,\beta)
        \ee
where $F(t,\beta)$ and $I(t,\beta)$ are defined with the corresponding $\rho^{1/4}$ insertions ~\cite{maldacena2016} with respect to the $\beta = 0$ analogues. 
As a consequence, it is only necessary to compute analytically $C(t,\beta)$ for a Hermitian $q = 2$ SYK model at finite temperature.

In order to proceed, we choose a basis in which $H$ is diagonal,
\be
H = i \sum_{k=1}^{N/2} \lambda_{k} \wchi_{2k-1} \wchi_{2k},
\ee
where $\wchi_i =S_{ik} \chi_k$, with $S_{ik}$ an orthogonal matrix, and $\lambda_k$ are the eigenvalues of the antisymmetric real couplings $J_{ij}$. 
In this new basis, the OTOC is given by
\be
F(t,\beta) &=&       \Tr \rho^{\frac 14} e^{iHt} \wchi_k  e^{-iHt} \rho^{\frac 14}
\wchi_m\rho^{\frac 14}
e^{iHt}\wchi_p e^{-iHt}\rho^{\frac 14} \wchi_q \nn\\
&&\times \left \langle \int dS S^T_{1k} S^T_{2m}  S^T_{1p} S^T_{2q} \right \rangle.
\ee
It is now necessary to carry out the averages over the orthogonal matrices $S_{ij}$. This is possible by employing the relation~\cite{ullah1963}
  \be
\langle S_{1k} S_{2m} S_{1p} S_{2q}\rangle=
\frac {N+1}{\mathcal{N}}\delta_{kp}\delta_{mq}
- \frac 1{\mathcal{N}} (\delta_{km}\delta_{qp}+\delta_{kq}\delta_{mp} ),
\nn\\
\ee
where ${\cal N} = N(N-1)(N+2)$, which results in three different contributions to  $F(t,\beta)$.
The first two are equal because of the reflection symmetry of the spectrum. The OTOC can be further simplified using $\wchi_k^2 = 1/2$ after commuting the four Majorana operators through the evolution operator and the density matrix.
The resulting sums are expressed in terms of $\lambda_k$ by using the properties of the trace over the many-body states. 
The final finite-$N$ result for 
$C(t,\beta)$ Eq.~(\ref{eq:ctfull}) is
\begin{align}
	F(t,\beta) =&   -\frac{N+1}{\mathcal{N}} \left(\sum_{k}^{N/2} \frac{1}{\cosh  \frac{ \beta\lambda_k}{2} } \right)^2   + \frac{N+2}{\mathcal{N}} \sum_{k}^{N/2} \frac{1}{\cosh^2  \frac{ \beta\lambda_k}{2} }    \notag\\
	& -\frac{2}{\mathcal{N}} \sum_k^{N/2}  \frac{\cos 2\lambda_k t}{\cosh  \frac{ \beta\lambda_k}{2}}  
	-\frac{2}{\mathcal{N}} \mathrm{Re}\Bigg[\! \left( \sum_{k}^{N/2}  \frac{\cos \lambda_k(t +i\frac{\beta}{4})}{\cosh  \frac{ \beta\lambda_k}{2}} \right)^2 \notag \\ & - \sum_{k}^{N/2} \frac{\cos 2\lambda_k(t +i\frac{\beta}{4})}{2\cosh^2  \frac{ \beta\lambda_k}{2}} \Bigg],   
	\label{eq:ctsum}\\
	I(t,\beta) =&  -\frac{1}{\mathcal{N}}  \sum_k^{N/2}  \frac{1}{\cosh  \frac{ \beta\lambda_k}{2}}  +  \frac{N+1}{\mathcal{N}}   \left( \sum_{k}^{N/2} \frac{\cosh \frac{\beta\lambda_k}{4} }{\cosh  \frac{ \beta\lambda_k}{2}} \right)^2   \notag\\
	&+\frac{N+2}{\mathcal{N}}   \sum_{k}^{N/2}  \frac{ \sinh^2 \frac{\beta \lambda_k}{4}  }{\cosh^2 \frac{ \beta\lambda_k}{2}}    +\frac{1}{\mathcal{N}} \left( \sum_{k}^{N/2} \frac{\cos \lambda_k t }{\cosh  \frac{ \beta\lambda_k}{2}} \right)^2   \notag \\ & 
	-\frac{1}{\mathcal{N}}   \left|\sum_{k}^{N/2} \frac{\cos \lambda_k (t - i\frac{\beta}{4}) }{\cosh  \frac{ \beta\lambda_k}{2}} \right|^2.     
	\label{eq:gtsum1}
\end{align} 

\section{Analytic expressions to orders $1/N$ and $1/N^2$}

\begin{figure}[t!]
	\includegraphics[width=\columnwidth]{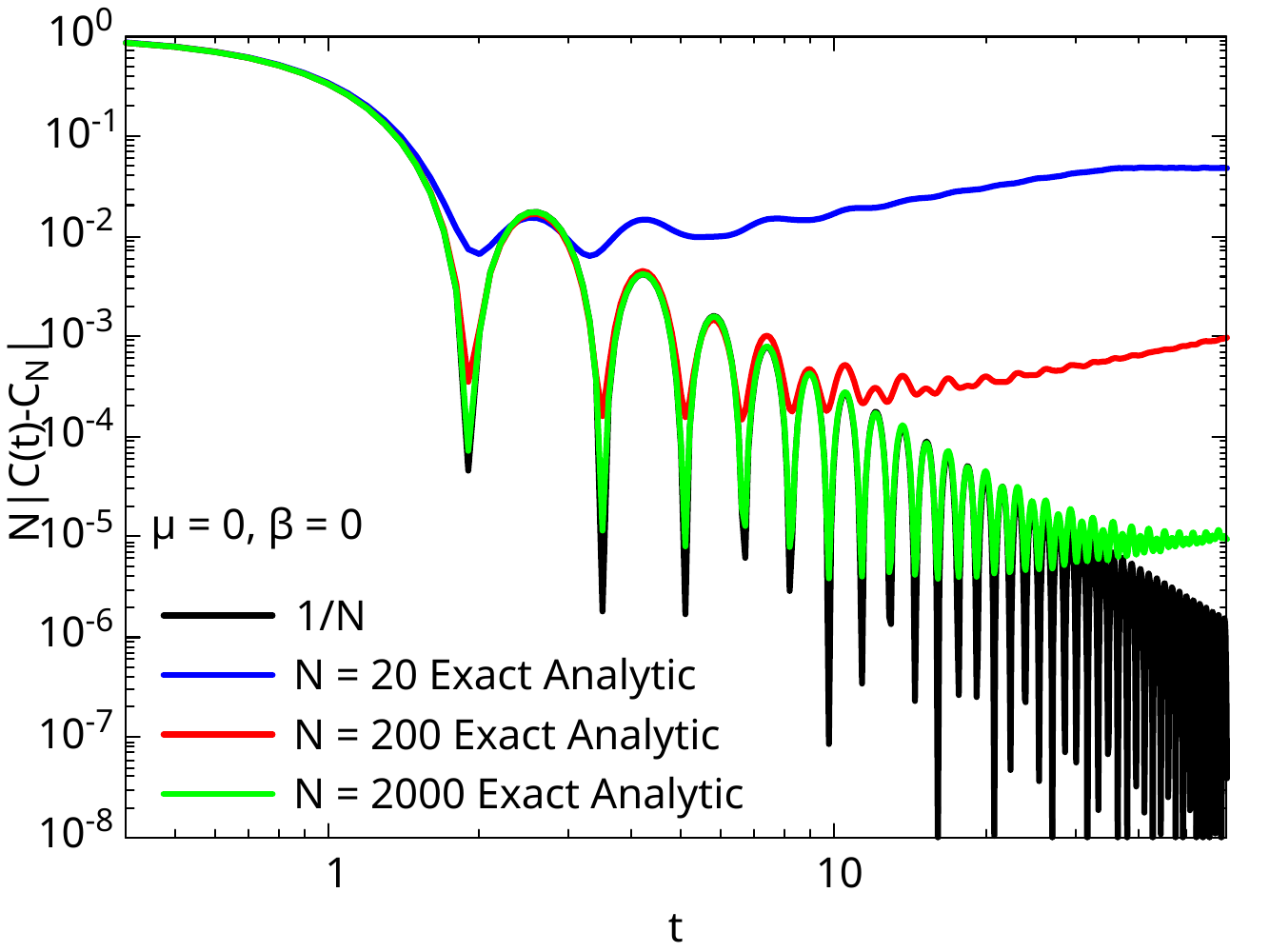}
	\includegraphics[width=\columnwidth]{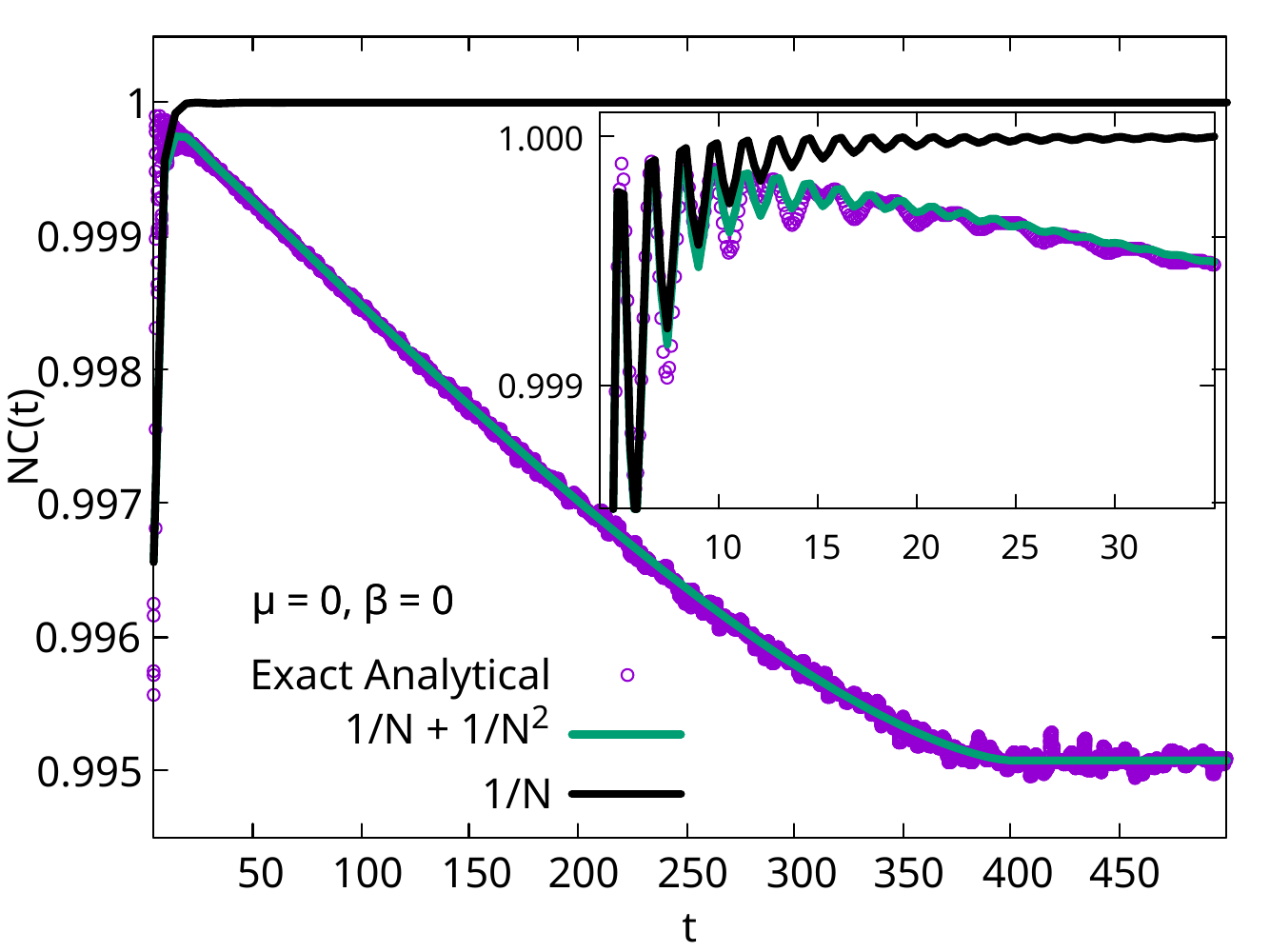}
	\caption{
	Time evolution of the quantum uncertainty $C(t)$ with $t$ in units of $J = 1$ for $\mu=\beta=0$. 
	Top: The blue ($N = 20$), red ($N = 200$), and green ($N = 2000$) curves correspond to $C(t)$ given in Eq.~(\ref{eq:ctsum}) and (\ref{eq:gtsum1}) and $C_N = (N+1)/[(N-1)(N+2)]$. They converge to the $1/N$ prediction Eq.~(\ref{eq:ctsemibeta0}), with $C_N = 1/N$, only for sufficiently large $N$. In both cases, $C_N$ is given by the time independent part of $C(t)$. 	
	Bottom: Comparison between the analytical exact $C(t)$, Eqs.~(\ref{eq:ctfull}), (\ref{eq:ctsum}) and  (\ref{eq:gtsum1}),  in terms of the eigenvalues of $J_{ij}$ (purple circles) with the {\it simple} analytic $C(t)$, Eq.  \eref{eq:ctn2}, including up to $1/N^2$ corrections for $N=200$ on a linear scale. The black curve stands for the $1/N$ result, Eq.~(\ref{eq:ctsemibeta0}), while the green curve stands for the $1/N+1/N^2$ result, Eqs.~(\ref{eq:ctn2}) and (\ref{eq:kct}).
     The agreement between of analytical exact and the $1/N+1/N^2$ results is excellent for all timescales. Deviations from the $1/N$ expression occur at $t\sim \sqrt{N}$. The Heisenberg time, signaling saturation, is at $t \sim 2 N$.
}
	\label{fig:compsum}
\end{figure}

The above expressions simplify by keeping only the leading $1/N$ correction,
replacing the sums with integrals and noticing that the spectral density of the eigenvalues $\lambda_k$ is $\rho(\lambda)=(1 /\pi) \sqrt{1-\lambda^2/4}$~\cite{mehta2004}. For simplicity, we henceforth restrict ourselves to $\beta = 0$. Considering only the leading $1/N$ correction, $C(t,0)\equiv C(t)$ Eq.~(\ref{eq:ctfull}), with $F(t,0)$ and $I(t,0)$ given by Eqs.~(\ref{eq:ctsum}),~(\ref{eq:gtsum1}), simplifies to
\begin{align}
	C(t) = \frac 1N \Bigg[
	&\int_{-2}^{2}  d \lambda \rho(\lambda) 
	-\left(\int_{-2}^{2}d\lambda \rho(\lambda)\cos(\lambda t)\right)^2
        \Bigg]{\rm e}^{-2\mu t},
        \label{eq:ctsemi}
\end{align}
which can be evaluated explicitly,
\be \label{eq:ctsemibeta0}
C(t) =\frac 1{N} \left(1- \frac{J_1^2(2t)}{t^2}\right){\mathrm e}^{-2\mu t},
\ee
where $J_1(t)$ is a Bessel function of first kind and time is measured in units of $J = 1$. The upper plot in Fig.~\ref{fig:compsum} demonstrates the convergence of the numerical results of $C(t)$ obtained from Eqs.~(\ref{eq:ctsum}) and (\ref{eq:gtsum1}) to the analytic prediction Eq.~(\ref{eq:ctsemibeta0}). 
Analogous analytic results for finite $\beta$ are rather cumbersome. In Appendix~\ref{app:ft}, we have checked that finite temperature effects do not change qualitatively the dynamics, at least for sufficiently long times.
A consequence of Eq.~(\ref{eq:ctsemibeta0}) is that, for $t \ll 1$, $C(t)$ grows quadratically as $t^2$. For $t \gg 1$, the system approaches a steady state in a power-law fashion with superimposed oscillations. 
The effect of the bath is an overall exponential decay of $C(t)$ that, as expected, suppresses quantum scrambling. For a sufficiently weak coupling, $\mu \ll 1$, the system approaches the steady state in two stages: first, the mentioned power-law, and only for longer times $\sim 1/\mu$ the exponential suppression of $C(t)$. 

We now proceed with the analytical evaluation of $1/N^2$ terms in $C(t)$ from Eqs.~(\ref{eq:ctsum}),~(\ref{eq:gtsum1}). This requires the inclusion of the two-point correlations of the eigenvalues of the coupling matrix $J_{ij} = -J_{ji}$, including self-correlations which can be expressed in terms of the spectral form factor.
A simple calculation shows that,
\be\label{eq:ctn2}
  C(t) &=&\frac{N^2}{\mathcal{N}}\left[1- \frac{J_1^2(2t)}{t^2}-K_c(t)\right .
\\
   && \left . +\frac{1}{N}\left(1-\frac{2J_1(2t)\theta(2N-t) +J_1(4t)}{2t}\right)\right]{\mathrm e}^{-2\mu t},\nn
\ee  
where we have reinstated $\cal N$ in order to account for additional $1/N^2$ effects, $\theta$ is the Heaviside step function, and
\be
N^2 K_c(t)= \langle |\sum_{k=1}^N \exp(i t \lambda_k)|^2\rangle-|\langle \sum_{k=1}^N \exp(i t \lambda_k)\rangle|^2
\ee
is the connected spectral form factor (SFF).

In the large-$N$ limit one can obtain \cite{hikami1997,forrester2021} an explicit analytic expression for $K_c(t)$ as follows.
Since $J_{ij}$ is an antisymmetric random matrix (class D) whose joint eigenvalue distribution ~\cite{mehta2004,gross2017} coincides with that of  the chiral Gaussian Unitary ensemble with topological number $\nu =-1/2$, its spectral density is a semicircle with
bulk two-point correlations given by the Gaussian Unitary Ensemble. For a non-uniform
spectral density, the spectral form factor is in general given by
\be
K_c(t)= \int dx \rho(x) K_{\rm RMT} (t/ \rho(x)),
\ee
where $K_{\rm RMT}$ is the universal RMT SFF for the appropriate ensemble.
This integral is evaluated as,
\begin{equation} \label{eq:kct}
K_c(t)=\begin{cases}
	\frac{2}{\pi N}\left (\frac{t}{2N}\sqrt{1-\frac{t^2}{4N^2}}+\arcsin(\frac{t}{2N})\right),\ t < 2N	\\
	\frac1N,\ t \geq 2N
\end{cases}
\!\!\!.
\end{equation}
For times $t \gtrsim \sqrt{N}$, $K_c(t) \sim 2t/\pi N$ and then gradually saturates to $1/N$ as the Heisenberg time $t \sim 2N$ is approached. A comment is in order. This SFF is computed from the eigenvalues of the couplings $J_{ij}$, so in line with previous results \cite{lea2019,torresherrera2017a}, it can be interpreted as the one obtained from the single-particle quantum chaotic motion. Therefore, we do not observe an exponential ramp as the one claimed in Refs.~\cite{winer2020,liao2020,legramandi2024}.
We stress that since the spectral density is not constant, the time dependence of Eq.~(\ref{eq:kct}) has clear deviations from linear behavior before the Heisenberg time, which translates into similar deviations in $C(t)$.
Interestingly, in the range of times of interest, $2N > t > \sqrt{N}$, a simple inspection of Eq.~(\ref{eq:ctn2}) reveals that $C(t)$ is controlled by $K_c(t)$ despite it being of the order $1/N^2$ and therefore subleading in the $1/N$ expansion. The reason for that is that the time dependence of the leading $1/N$ terms approaches zero in a power-law fashion so that for $t > \sqrt{N}$ they become smaller than $K_c(t)$ which increases linearly in time. Crucially, the sign of the SFF term is negative, which leads to a local maximum in the OTOC not related to the small oscillations.
Therefore, the OTOC decreases with time for $t > \sqrt{N}$, which corresponds to a decrease of quantum scrambling due to the discreteness of the spectrum captured by the SFF. 

The results depicted in the bottom panel of Fig.~\ref{fig:compsum} fully confirm this picture. 
The analytical expression for $C(t)$ Eq.~(\ref{eq:ctn2}), which includes $1/N$ and $1/N^2$ corrections, agrees at all times scales with the exact $C(t)$ Eqs.~(\ref{eq:ctfull}), (\ref{eq:ctsum}), (\ref{eq:gtsum1}), at $\beta = 0$, in terms of the eigenvalues of $J_{ij}$, which must be computed numerically. 
Substantial deviations are observed for long times if only the leading $1/N$ correction Eq.~(\ref{eq:ctsemibeta0}) is taken into account. The effect of the bath is to suppress scrambling exponentially.

We have thus found that, for $t \gtrsim \sqrt{N}$, $C(t)$ is controlled by the SFF of the
random couplings $J_{ij}$, which has the expected linear behavior observed in quantum chaotic systems for times $t > \sqrt{N}$ but not too close to the Heisenberg time.
This does not mean that the $q=2$ SYK is many-body quantum chaotic. While single-particle observables may show quantum chaos, the many-body dynamics is still integrable because the Majoranas are not interacting. For instance, the level statistics is Poisson. Moreover, the timescales involved are polynomial in $N$ which is typical of integrable systems. 
The results of Appendix~\ref{app:ft} show that finite $\beta$ effects do not change qualitatively this picture. Before we embark on the comparison with numerical results, we comment on the differences and similarities between the $C(t)$ of a quantum chaotic system and that of the integrable SYK model. Starting with the similarities, the perturbative quadratic growth and the power-law decay have also been observed in quantum chaotic systems \cite{lea2019,torresherrera2017,torresherrera2017a,garcia2018a}. Moreover, the observed non-monotonicity is also expected in quantum chaotic systems because it is a general consequence of the discreteness of the spectrum. Finally, at least for short times \cite{garcia2024}, the effect of the bath induces a similar decay in both cases. However, there are also differences, for instance, the OTOC for the integrable SYK model experiences neither the exponential growth around the Ehrenfest time nor the exponential decay characterized by the Ruelle-Pollicott resonances \cite{jalabert2018,garciamata2023} as the system approaches thermalization, which are features in the quantum chaotic case. Finally, the saturation time is $\sim \exp(N/2)$ for quantum chaotic many-body systems while it is linear $\sim N$ for the integrable SYK model.
	
\begin{figure}[t!]
	\includegraphics[width=\columnwidth]{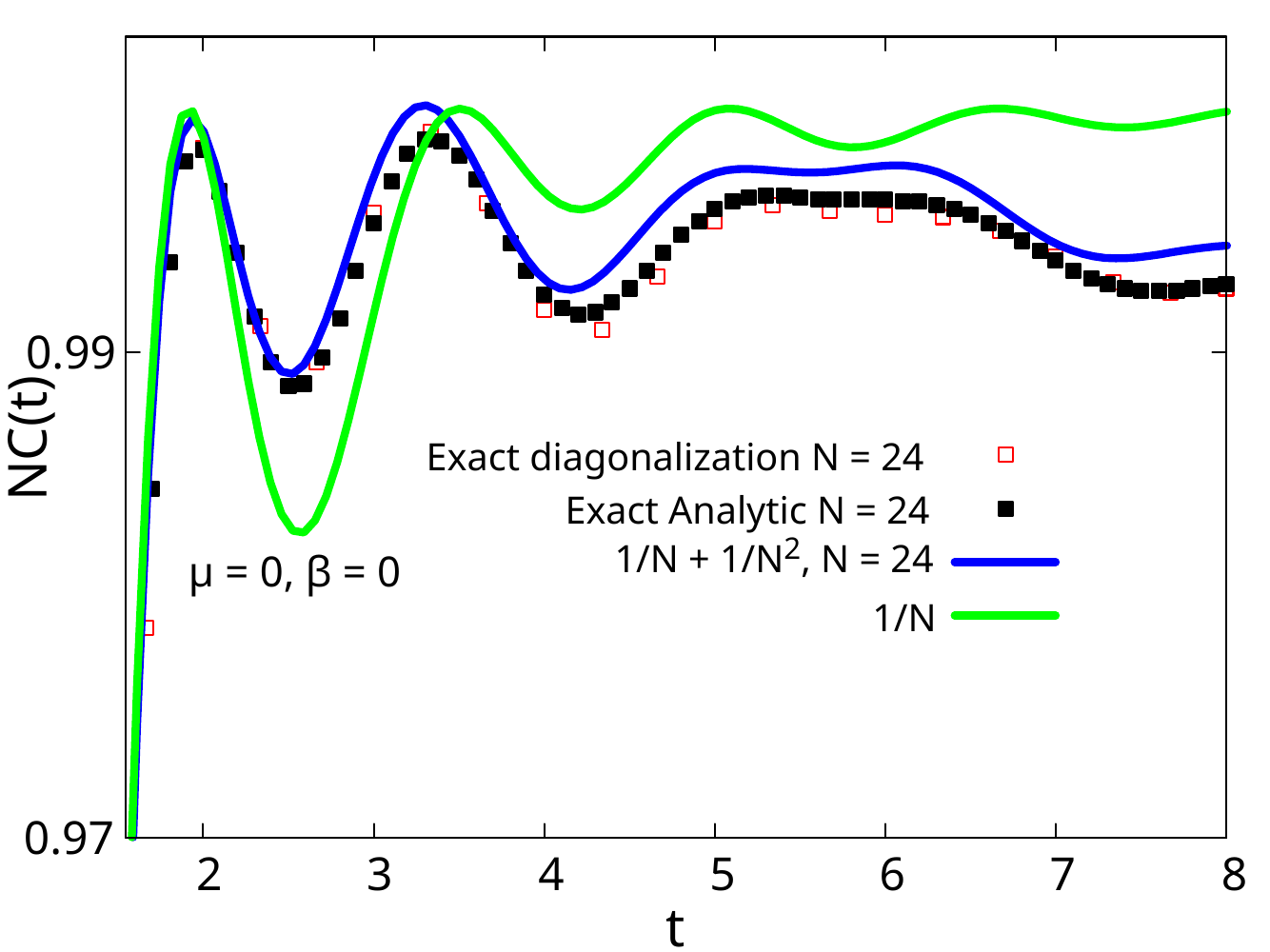} 
	\includegraphics[width=\columnwidth]{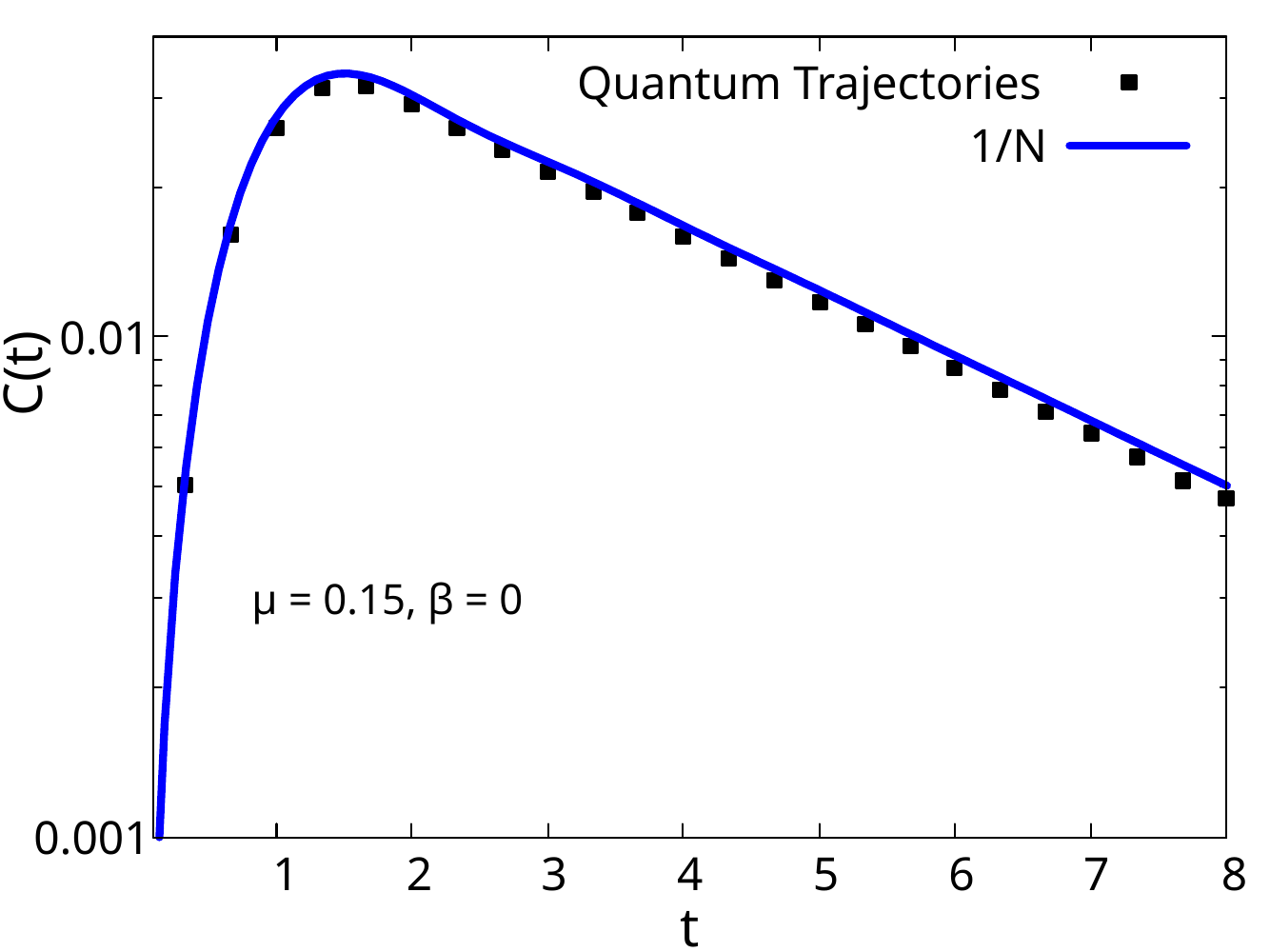}
	\caption{
	Top: $C(t)$ with $t$ in units of $J = 1$, for $\mu = \beta = 0$. $C(t)$ Eq.~(\ref{eq:ctfull}) is computed using exact diagonalization for $N = 24$ (red) with $\sim 10^6$ disorder realizations.
 	The exact diagonalization result shows almost no difference with respect to the exact analytic result (black), Eqs.~(\ref{eq:ctsum}) and (\ref{eq:gtsum1}), and and also an excellent agreement with the $1/N+1/N^2$ {\it simple} expression (blue) given in  Eq.~(\ref{eq:ctn2}). Substantial deviations are observed if only the leading $1/N$ correction (green) Eq.~(\ref{eq:ctsemibeta0}) is kept. 
 	We are neglecting terms $1/N^2$ in $NC(t)$. Since $N = 24$ and, at least to lower order, prefactors are of the order one, differences between ED and the analytic $1/N+1/N^2$  expression (blue curve) should be $\sim 1/24^2 \approx 0.0017$, is  fully consistent with the observed deviation.
	Bottom: $C(t)$ for $\mu = 0.15$, $N = 18$ and $\beta = 0$. We combine exact diagonalization with the quantum trajectory method~\cite{molmer93,dum1992}. 
	For each point in time, we use $200$ quantum trajectories, a time step of $dt = 0.01$, and at least $5\times10^5$ disorder realizations. 
	The exponential approach to the steady state for $\mu\ne 0$ is correctly reproduced by the quantum trajectory method and we find good agreement with the $1/N$ semiclassical result.
	}
	\label{fig:otoc}
\end{figure}

\section{Exact numerical results}

Previously, we carried out a comparison between {\it simple} analytical expressions depending on elementary functions, Eqs.~(\ref{eq:ctsemibeta0}) and (\ref{eq:ctn2}), and {\it full} analytical expressions, Eqs.~(\ref{eq:ctsum}) and (\ref{eq:gtsum1}), that still require the diagonalization of a single-particle random matrix $J_{ij}$. We now compare those analytical results with a numerical calculation of $C(t)$ by exact diagonalization 
combined with the quantum trajectory method~\cite{dalibard1992,molmer93,dum1992} when the bath is turned on ($\mu \neq 0$). The fermionic quantum trajectory method is described in Appendix~\ref{app:qtrajectory}. The need of a double average over quantum trajectories and disorder realizations for $\mu > 0$ limits the sizes for which we can make a detailed comparison to $N \lesssim 20$. 

The results in Fig.~\ref{fig:otoc} for $\mu = 0$ show 
an excellent agreement between the compact analytical $C(t)$ Eq.~(\ref{eq:ctn2}), including $1/N+1/N^2$ contributions, and the exact diagonalization result. The small shift upward of the former is of order $1/N^3$ and therefore consistent with neglected terms in the expansion. Differences between the exact analytical $C(t)$ Eqs.~(\ref{eq:ctsum}), (\ref{eq:gtsum1}) and exact diagonalization results are barely noticeable.  We stress that the comparison is parameter free and that the inclusion of $1/N^2$ corrections is essential for the observed level of agreement. 
For a finite $\mu$ (bottom plot), the numerical $C(t)$, using $200$ quantum trajectories, reproduces correctly the expected ${\rm e}^{-2 \mu t}$ decay due to the bath, a feature that we derive analytically in Appendix \ref{app:qtrajectory}.
Subleading features, like oscillations, are difficult to reproduce numerically because it would require a much larger number of disorder realizations. This illustrates the importance of obtaining analytic results to describe quantitatively the different stages of the quantum dynamics.

\section{Conclusions}

We studied information scrambling and the effect of Markovian dissipation in an integrable SYK model through an analytical calculation of the growth of quantum uncertainty characterized by OTOCs at all times scales. 
In the absence of dissipation, the asymptotic approach to the steady state is power-law with superimposed oscillations. 
At $t \sim \sqrt{N}$, the overall growth of scrambling stops and starts to decrease linearly in time. This unexpected change of trend has its origin in a subleading $1/N^2$ correction to the OTOC that becomes dominant in this region. We show analytically that this correction is nothing but the SFF of the random couplings $J_{ij}$.
It becomes dominant for $t \gtrsim \sqrt{N}$ because the SFF is linear in time, whereas the rest of the time-dependent terms in the OTOC tend to zero in a power-law fashion. For $t \approx 2N$ (the Heisenberg time), the linear decrease of the uncertainty terminates, 
and the uncertainty reaches saturation though we still observe small oscillatory contributions whose amplitude decreases as a power law in time.
The effect of the Markovian environment is an exponential decay of the growth of quantum uncertainty at a timescale inversely proportional to the coupling to the bath, so it will eventually dominate the approach to saturation. 
We believe that features like a power-law decay approach to saturation and a local maximum of the quantum uncertainty due to the discreteness of the spectrum are generic features in quantum many-body integrable systems. It would be interesting to explore the dynamics of the present SYK model employing more general jump operators so that the vectorized Liouvillian has random quartic terms in order to study whether the environment can induce quantum chaotic features in the dynamics. 

\acknowledgments{
AMGG thanks Klaus M{\o}lmer for illuminating correspondence.
 AMGG, JPZ, and CL acknowledge support from the National Natural Science Foundation of China (NSFC):
Individual Grant No.\ 12374138, Research Fund for
International Senior Scientists No.\ 12350710180, and National Key R$\&$D Program of China (Project ID: 2019YFA0308603). 
AMGG acknowledges support from a Shanghai talent program.
LS was supported by a Research Fellowship from the Royal Commission for the Exhibition of 1851.
JJMV is supported in part by US DOE Grant No.\ DE-FAG-88FR40388. 
}

\renewcommand{\appendixname}{APPENDIX}
\appendix 

\section{\uppercase{Finite-temperature results}}
\label{app:ft}

In this appendix we obtain jump operators such that the system relaxes
 to the thermal state~\cite{prosen2008,breuer2002} and show that
also in this case the coupling to the bath factorizes from the OTOC. We then show that finite temperature does not qualitatively change the behaviour of $C(t)$.

\subsection{Finite-temperature steady state}
By an orthogonal transformation of the Majorana fermions, $\tilde{\chi}_i = S_{ij} \chi_j$,
the $q = 2$ Hamiltonian of the SYK model in Eq.~\eref{eq:densitymsyk2} can be rewritten as~\cite{cotler2016}
\begin{equation}\label{eq:sykd}
	H = i\sum_{i<j}^{N} J_{ij} \chi_i \chi_j 
	= i\sum_{k=1}^{N/2} \lambda_k \tilde{\chi}_{2k-1} \tilde{\chi}_{2k},
\end{equation}
where $i\lambda_k$ are the eigenvalues of $J_{ij}$. In terms of Dirac fermions ${c}_k^\dagger = (\tilde{\chi}_{2k-1} - i\tilde{\chi}_{2k})/\sqrt 2$, the Hamiltonian reads
\begin{equation}
    H = \sum_{k=1}^{N/2} \lambda_k \left({c}^\dagger_k {c}_k - \frac{1}{2} \right) \equiv \sum_{k=1}^{N/2} H_k.
\end{equation}
 We now show that the jump operators
\begin{align}\label{eq:jumpt}
   L_k^{(1)} = \sqrt{\mu(1 - f(\lambda_k))}{c}_k,\quad
   & L_{k}^{(2)} = \sqrt{\mu f(\lambda_k)}{c}_k^{\dagger},
   \end{align}
where $f(\lambda)=1/(1 +\exp({\beta\lambda}))$ is the Fermi-Dirac distribution,
lead to the steady state density matrix $\rho = e^{-\beta H}/Z$.
In terms of these jump operators, the Lindblad equation becomes
\begin{align}
  \mathcal{L}_\beta[\rho] = &-i[H,\rho] +\mu\sum_{k=1}^{N/2}\Big[
          (1-f(\lambda_k))(2{c}_k \rho {c}_k^{\dagger} - \{{c}_k^{\dagger}{c}_k,\rho\})
	\nn \\
	&+ f(\lambda_k)(2{c}_k^{\dagger}\rho {c}_k - \{{c}_k {c}_k^{\dagger},\rho\}) \Big]. 
\end{align}
Since the $H_k$ commute among themselves and also commute with $c_{l\ne k}$ and
$c_{l\ne k}^\dagger$,
we only have to consider the commutation of the factor $\exp(-\beta H_k)$ with $c_k$ and
$c_k^\dagger$. Using that $c_k\exp(-\beta H_k) c_k^\dagger = \exp(-\beta \lambda_k/2 )c_kc_k^\dagger$,
etc., one easily sees that $\mathcal{L}_\beta(\exp(-\beta H))=0$.

\subsection{Factorization of the $\mu$ dependence}
The time evolution of $c_p$ and $c_p^\dagger$ is given by the adjoint Lindblad operator~\cite{breuer2002},
\begin{align}
	& \frac{d c_p(t)}{dt}
	= \mathcal{L}^\dagger[{c_p}(t)]
	= +i[H,c_p(t)] 
	\notag\\
	&+\mu\sum_k^{N/2} \Big[(1-f(\lambda_k))(-2{c}_k^{\dagger} c_p(t) {c}_k
          - \{{c}_k^{\dagger}{c}_k,c_p(t)\}) 
	\notag\\
	&+ f(\lambda_k)(-2{c}_k c_p(t) {c}_k^{\dagger}  - \{{c}_k {c}_k^{\dagger},c_p(t)\})\Big], 
\end{align}
and the same equation for the evolution of $c_p^\dagger$.
From the anti-commutation relations of the creation and annihilation operator we find
\be
 \frac{d c_p(t)}{dt}
	= \mathcal{L}^\dagger[{c_p}(t)]
	= i[H,c_p(t)] -\mu c_p(t),
        \ee
        and the same evolution equation for $c_p^\dagger(t)$. Since the Majorana operators are linear
        combinations of $c_p$ and $c_p^\dagger$ we thus find that
        \be
\left(        \frac d{dt} +\mu \right) \tilde{\chi}_k (t)= i [H, \tilde{\chi}_k(t)].
\label{chimaster}
\ee
Because the original Majoranas are linearly related to the $ \tilde{\chi}_k$ this equation
also holds for the $\chi_k$. This shows that the $\mu$ dependence also factorizes
at finite temperature.

\begin{figure}[t!]
	\includegraphics[width=\columnwidth]{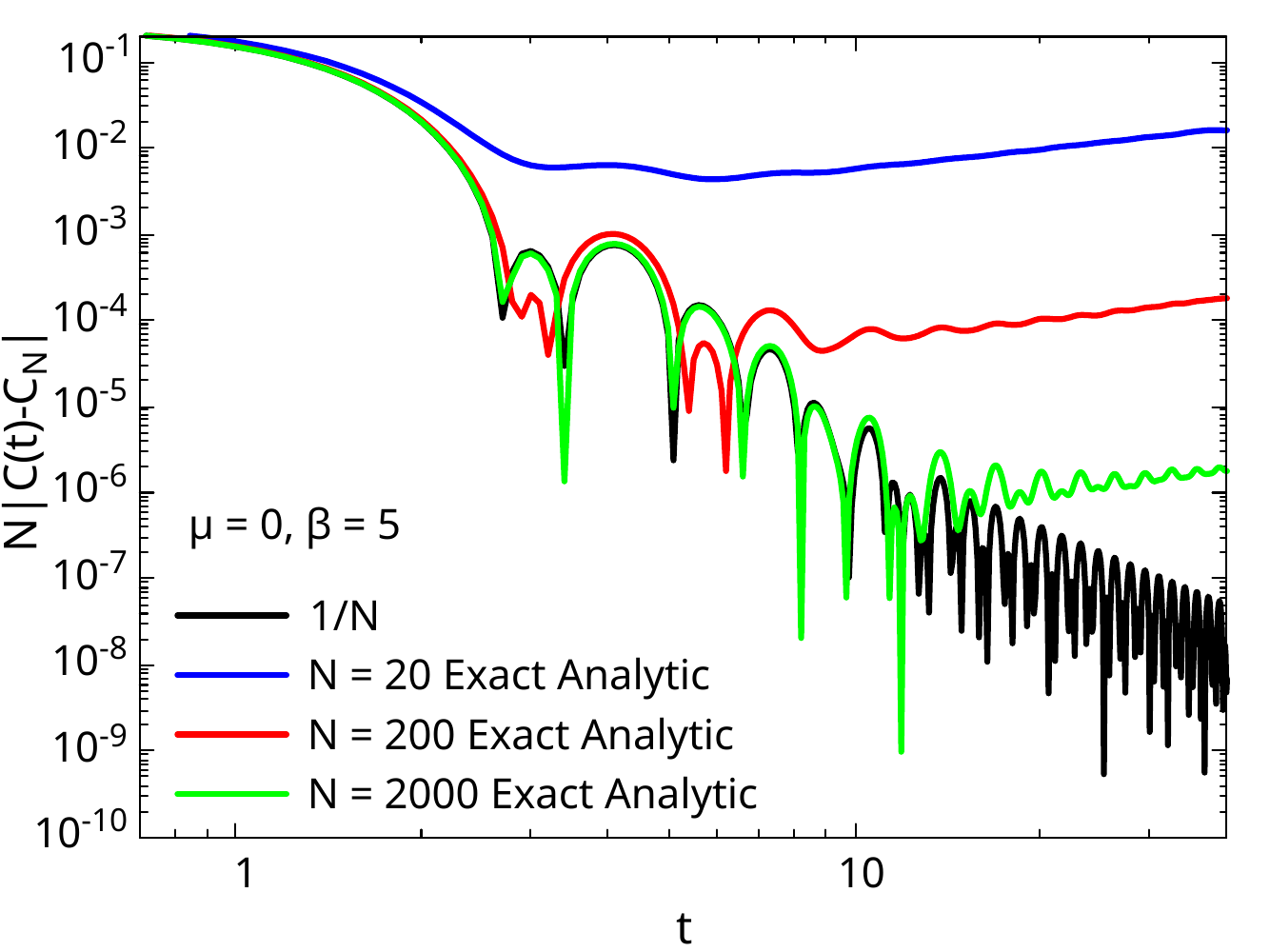}
	\includegraphics[width=\columnwidth]{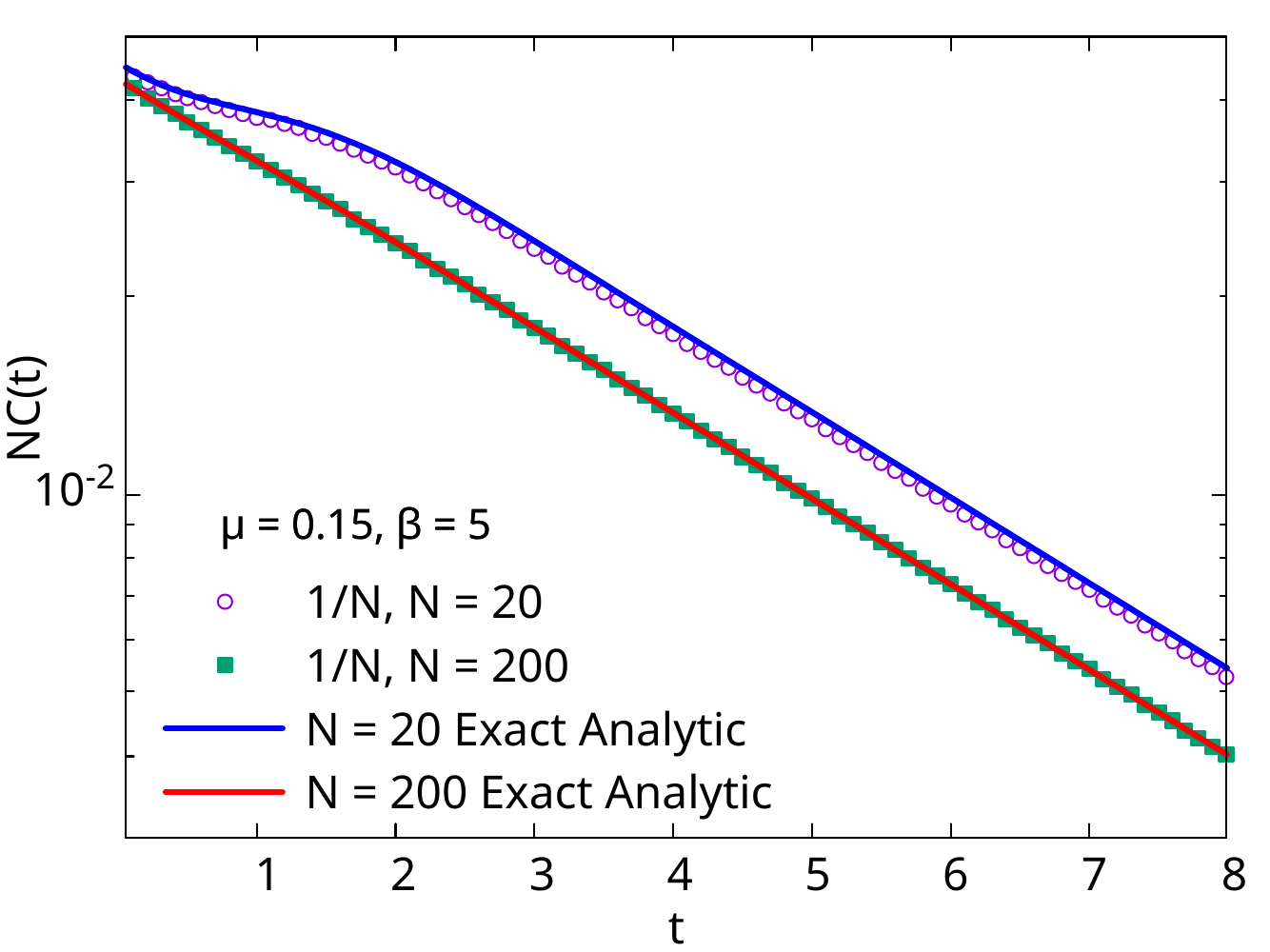}
	\caption{$C(t)$ with $t$ in units of $J = 1$ for $\beta=5$. 
		Top: $\mu = 0$. The black curve stands for the semiclassical expression Eq.~(\ref{eq:ctsemifinitebeta}) with $C_N$ given by the time independent part of the coefficient
		of $\exp[-\mu t]$ in Eq.~(\ref{eq:ctsemifinitebeta}). The blue ($N = 20$), red ($N = 200$), and green ($N = 2000$) curves stand for the exact analytic $C(t)$, Eqs.~(\ref{eq:ctsum}) and (\ref{eq:gtsum1}) with $C_N = 0.06, 0.044, 0.043$ for $N = 20, 200, 2000$, respectively. The initial approach to the steady state is still power-law, so thermal effects do not lead to qualitative changes. 
		Bottom: $\mu = 0.15, \beta = 5$. For $t \gtrsim 1/2\mu$ the decay is exponential and fully controlled by the Markovian bath.}
	\label{fig:finitebeta}
\end{figure}

\subsection{Dynamics of $C(t)$}
At finite $\beta$, the leading $1/N$ expression for $C(t)$ is [recall that at $\beta=0$ we have Eq.~(\ref{eq:ctsemi})]:
\begin{align}\label{eq:ctsemifinitebeta}
	&C(t) = 
	\frac{1}{2N}\Bigg[
	2\int_{-2}^{2}  \frac{d \lambda \rho(\lambda)}{\cosh  \frac{ \beta\lambda}{2}}
	+2\left(\int_{-2}^{2}\frac{d\lambda \rho(\lambda)\cos(\lambda t)}{\cosh \frac{ \beta\lambda}{2}}\right)^2
	\nn \\ &
	-N\left(\int_{-2}^{2}\frac{d\lambda \rho(\lambda)}{\cosh\frac{ \beta\lambda}{2}}\right)^2
	+ N  \left(\int_{-2}^{2}\frac{d\lambda \rho(\lambda)\cosh\frac{ \beta\lambda}{4}}{\cosh\frac{ \beta\lambda}{2}}\right)^2  
	\nn \\ &
	-3\left(\int_{-2}^{2}d\lambda \rho(\lambda)\frac{\cos(\lambda t)\cosh\frac{ \beta\lambda}{4}}{\cosh\frac{ \beta\lambda}{2}}\right)^2
	\nn \\ &
	+ \left(\int_{-2}^{2}d\lambda \rho(\lambda)\frac{\sin(\lambda t)\sinh\frac{ \beta\lambda}{4}}{\cosh\frac{ \beta\lambda}{2}}\right)^2
	\Bigg]{\rm e}^{-2\mu t}.
\end{align} 
The one-dimensional integrals in Eq.~(\ref{eq:ctsemifinitebeta}) cannot be evaluated exactly but an asymptotic expression valid in the limit $t \gg \beta$ with $\beta$ and $\mu$ fixed is available.
The most salient finite-$\beta$ effect in $C(t)$ is to induce, for intermediate times, an exponential decay $\sim e^{-\beta \pi t}$.
For longer times, the approach to the steady state is power-law $\sim 1/t^3$ as in the $\beta = 0$ case. At finite $\mu$, the asymptotic decay is exponential and $\beta$-independent, with oscillations still superimposed. Therefore, thermal effects are mostly quantitative not qualitative. This is confirmed by an explicit comparison, depicted in Fig.~{\ref{fig:finitebeta}, between the $1/N$ prediction Eq.~(\ref{eq:ctsemifinitebeta}) and the exact $C(t)$ Eqs.~(\ref{eq:ctsum}) and (\ref{eq:gtsum1}) for $\beta = 5$.

Regarding $1/N^2$ corrections, an analogous expression to Eq.~(\ref{eq:ctn2}) could also be worked out at finite $\beta$, although it would be rather cumbersome. Moreover, the results of Fig.~\ref{fig:finitebeta} show that thermal effects do not induce quantitative changes in $C(t)$.

\section{\uppercase{Fermionic Quantum Trajectory Method}}
\label{app:qtrajectory}

In this appendix, we first review the method of quantum trajectories
\cite{dalibard1992,molmer93,dum1992}, originally proposed in the context of bosonic systems and extend it to the fermionic systems
discussed in this paper. We consider an open quantum system where the system degrees of freedom consist of $N$ bosonic fields $a_i, i=1,\ldots,N$ with $[a_i,a_j^\dagger]=\delta_{ij}$. When coupled to a Markovian bath, the reduced density matrix $\rho$ for the system evolves under the following master equation
\be \label{eq:app_Lindblad}
\frac{d\rho}{dt} =- \ii [H, \rho] - \mu N \rho + \mu\sum_n L_n \rho L^\dagger_n\, ,
\ee
where $L_n$ are jump operators built out of the field operators $a_i$ and $a_i^\dagger$. Using the quantum regression theorem, the evolution equation for the Green's functions
$G_{ij}(t) = \langle a_i(t) a_j \rangle$ is given by the action of the adjoint Lindblad operator,
\begin{equation}\label{eq:bosonic}
	\frac{dG_{ij}}{dt} = \ii\langle [H, a_i](t) \, a_j \rangle - \mu N G_{ij}(t)  + \mu\sum_n \langle (L^\dagger_n a_i L_n)(t)\,a_j \rangle.
\end{equation}
The method of quantum trajectories states that this equation
is solved by
\begin{equation}
	G_{ij}(t) = \lim_{M \to\infty}\frac{1}{M} \sum_{q=1}^M \langle U^\dagger_{q} (t) a_i
	U_{q}(t) a_j \rangle,
\end{equation}
where each $U_{q}(t)$ is a unitary operator constructed in the following way: we first subdivide the time interval $[0,t]$ into equal segments of length $\delta t$. At each time step $k\delta t$, we generate a real number $\xi$ from the uniform distribution in $[0,1]$. Assuming the jump operators are unitary, the evolution operator from $k\delta t$ to $(k+1)\delta t$ is
\begin{equation}
	U_q(k \delta t) = \begin{cases}
		e^{-\ii H\delta t}, &\textrm{if $\xi > 1-e^{-\mu N\delta t/2}$} \\
		L_n, &\textrm{if $\xi < 1-e^{-\mu N\delta t/2}$}
	\end{cases}.
	\label{B3}
\end{equation}
Here, $n$ is an integer randomly sampled from the set $\{1, \ldots, N \}$ at each time step. The evolution operator $U_{q}(t)$ is then a product of these evolution operators.
\begin{equation}
	U_{q}(t) = U(t-\delta t) U(t-2\delta t) \cdots U(\delta t) U(0).
\end{equation}

In order to extend these results to fermonic systems such as the ones considered in the main text, a central issue is the anti-commutation relations of the fermionic field operators that replace the commutation relations assumed in the analysis above. When the jump operators and the fermion operators anti-commute, the quantum master equation (\ref{eq:app_Lindblad}) governing the evolution of the reduced density matrix can no longer be directly applied to the $n$-point function. Instead, as is shown in Appendix B of Ref.~\cite{PhysRevB.94.155142}, the adjoint Lindblad operator acts on the Green's function with an extra minus sign. More specifically, using the notations of the main text where $\chi_i, i=1,\ldots,N$ are $N$ Majorana fermions with anti-commutation relations $\{\chi_i,\chi_j\}=\delta_{ij}$, Eq.~(\ref{eq:bosonic}) should be replaced with ($ G_{ij}(t) = \langle \chi_i(t) \chi_j \rangle$)
\begin{equation}\label{eq:fermionic}
	\frac{dG_{ij}}{dt} = \ii\langle [H, \chi_i](t) \, \chi_j \rangle - \mu N G_{ij}(t)  - \mu\sum_n \langle (L^\dagger_n \chi_i L_n)(t)\,\chi_j \rangle.
\end{equation}
Note the minus sign in front of the last term. Since the rest of the equation is the same as the bosonic case considered above, we can account for this minus sign by replacing $L_n$ with $\tilde{L}_n = \gamma_c L_n$, where $\gamma_c$ is the chiral fermion operator. Using the fact that $\gamma^2_c=1$ and that $\gamma_c$ anti-commutes with all Majorana operators, we find
\begin{equation}
	\tilde L^\dagger_n \chi_i(t) \tilde L_n = L^\dagger_n \gamma_c \chi_i(t) \gamma_c L_n = - L^\dagger_n \chi_i(t) L_n.
\end{equation}
Therefore, replacing $L_n$ with $\tilde L_n$ converts the fermionic master equation to the bosonic one. Instead of Eq.~\eref{B3}, we use the following quantum trajectory evolution operator to obtain the solution of Eq.~(\ref{eq:fermionic}) ,
\begin{equation}
	\tilde U_q(k \delta t) = \begin{cases}
		e^{-\ii H\delta t}, &\textrm{if $\xi > 1-e^{-\mu N\delta t/2}$} \\
		\gamma_c L_n\equiv \sqrt 2 \gamma_c\chi_n , &\textrm{if $\xi < 1-e^{-\mu N\delta t/2}$}
	\end{cases}.
\end{equation}

The $\mu$ dependence of this time evolution can be worked out analytically when we consider a single time step. Since at every time step we have a probability $\mu N \delta t/2$ for a jump operator and a probability $(1- \mu N\delta t/2)$ for a Hamiltonian evolution, we have
\be \label{chiqt}
&&\hspace*{-0.5cm}\frac 1M \sum_{q=1}^M \tilde U^\dagger_{q}(\delta t) \chi_i (t)
\tilde U_{q}(\delta t)
= \left(1- \frac{\mu N\delta t}{2}\right)  e^{iH\delta t} \chi_i(t)  e^{-iH\delta t} \nn \\ &+& \frac{\mu N \delta t}{2}\frac{(N-2)}{N}\chi_i(t) = e^{iH\delta t} \chi_i(t)  e^{-iH\delta t} - \mu\delta t \chi_i(t),
\ee
where we have used $ \tilde L_n \chi_i \tilde L_n = \chi_i$
for $i \ne n$ and $ \tilde L_i \chi_i \tilde L_i = -\chi_i$.
For $q=2$ and to linear order in $\delta t$, this implies that $\chi_i$ satisfies the adjoint master equation Eq.~(\ref{chimaster}), implying that for $\mu t\ll N$ the $\mu$ dependence factorizes as
\be
G_{ij}(t) = e^{-\mu t } \langle e^{iHt} \chi_i e^{-iHt} \chi_j\rangle.
\ee
We emphasize that the additional minus sign in the  master equation was essential to obtain the correct $\mu$ dependence~\cite{garcia2022e}. Had we not included it, we would instead obtain a factor $e^{-\mu N t}$, which is what one expects for a bosonic system.

The procedure for the four-point function considered in the main text follows by replacing both time evolutions by the right-hand side of Eq.~\eref{chiqt} using \emph{independent} trajectories. From the argument above, it is clear that the $\mu$ dependence factorizes for the four-point function as well.

\bibliography{8_29otocsyk2025.bbl}

\end{document}